# Enhanced Thermoelectric Efficiency via Orthogonal Electrical and Thermal Conductances in Phosphorene

Ruixiang Fei,[1] Alireza Faghaninia,[2] Ryan Soklaski,[1] Jia-An Yan,[3] Cynthia Lo,[2] and Li Yang[1*]

[1] Department of Physics, Washington University, St Louis, MO, USA

[2] Department of Energy, Environmental & Chemical Engineering, Washington University, St Louis, MO, USA

[3] Department of Physics, Astronomy and Geosciences, Towson University, Towson, MD, USA

**Abstract**

Thermoelectric devices that utilize the Seebeck effect convert heat flow into electrical energy and are highly desirable for the development of portable, solid state, passively-powered electronic systems. The conversion efficiencies of such devices are quantified by the dimensionless thermoelectric figure of merit (*ZT*), which is proportional to the ratio of a device's electrical conductance to its thermal conductance. High *ZT* (>2) has been achieved in materials via all-scale hierarchical architecturing. This efficiency holds at high temperatures (700K~900K) but quickly diminishes at lower temperatures.  In this paper, a recently-fabricated two-dimensional (2D) semiconductor called phosphorene (monolayer black phosphorus) is assessed for its thermoelectric capabilities. First-principles and model calculations reveal that phosphorene possesses spatially-anisotropic electrical and thermal conductances. The prominent electrical and thermal conducting directions are orthogonal to one another, enhancing the ratio of these conductances. As a result, *ZT* can reach 2.5 (the criterion for commercial deployment) along the armchair direction of phosphorene at T=500K and is greater than 1 even at room temperature given moderate doping (~2 x $10^{16}$ $m^{-2}$). Ultimately, phosphorene stands out as an environmentally sound thermoelectric material with unprecedented qualities: intrinsically, it is a mechanically flexible material that converts heat energy with high efficiency at low temperatures (~ 300K) – one whose performance does not require any sophisticated engineering techniques.

**Corresponding Author**: * lyang@physics.wustl.edu

The task of designing practical and efficient thermoelectric materials is a decades-old problem that has spurred diverse and creative material engineering efforts for its purpose.[1,2] The efficiency of thermoelectric materials is measured by its figure of merit, $ZT = S^2 \frac{\sigma}{\kappa} T$, where $S$ is the Seebeck coefficient, σ is the electrical conductivity, κ is the thermal conductivity, and $T$ is the temperature. The forefront of all-scale electronic and atomistic structural engineering techniques have served to enhance the $\sigma/\kappa$ ratios of third-generation thermoelectric materials, such as Cubic AgPb$_m$SbTe$_{2+m}$, PbTe-SrTe and In$_4$Se$_{3-\delta}$,[3-5] achieving $ZT$ values near 2 (around a 15% conversion efficiency) within a temperature range of 700 K to 900K.[6] Alternatively, promising simple structures exhibit intrinsically low thermal conductances, without requiring sophisticated structural engineering, and achieve excellent thermoelectric performances.[7,8] For example, SnSe crystal can reach a $ZT$ of 2.6 at T= 923K,[8] although this value falls quickly for lower temperatures.

Graphene-inspired 2D materials also possess unusual thermal properties in addition along with optical and mechanical advantages [9-13], yet little progress has been made toward harnessing them for thermoelectric applications. Recently, a newly-fabricated direct band gap semiconductor, few-layer black phosphorous (phosphorene),[14-17] stands out as a possible intrinsically-efficient thermoelectric material. Along with possessing promising free-carrier mobility, it exhibits highly anisotropic electrical and optical responses [18-20]. It is natural to check if the thermal conductance is also anisotropic. If this is the case, it is desirable for the preferred direction of electrical conductance in phosphorene to coincide with one of poor thermal conductance; such an alignment would significantly enhance phosphorene's thermoelectric performance, as shown in Figure 1.

In this paper, first-principles simulations and model calculations of monolayer phosphorene are used to study its electrical conductance, thermal conductance, Seebeck coefficient, and thermoelectric figure of merit ($ZT$). Excitingly, the electrical and thermal conductances are shown not only to be anisotropic, but to exhibit respectively orthogonal preferred conducting directions, suggesting that a large $\sigma/\kappa$ ratio can be achieved in this material. These results are illustrated in Figure 1. Additionally, the Seebeck coefficient is calculated using the Boltzmann transport equation (BTE) [21]. It is ultimately found that $ZT$ can reach 1 at room temperature and

2.5 at T=500K using only moderate doping (~2 x $10^{16}$ m$^{-2}$). Therefore, monolayer phosphorene stands out as a mechanically-flexible thermoelectric material that can operate at room temperature, and does not require extensive structural engineering.

*Anisotropic electrical conductance:* The atomic structure of monolayer phosphorene is presented in Figure 2 (a). Each phosphorous atom is covalently bonded with its three neighboring phosphorous atoms to form a puckered 2D honeycomb structure. The electronic band structure of suspended monolayer phosphorene is presented in Figure 2 (b). The Γ point hosts a 2.0 eV direct band gap. The band dispersion about the Γ point is highly anisotropic; both the valence and conduction bands are significantly dispersive along the armchair lattice direction (Γ-Y) but are nearly flat along the perpendicular zigzag lattice direction (Γ-X). Accordingly, the effective masses of the free carriers in phosphorene, and thus its electrical conductance, are all highly anisotropic. Figures 2 (c) and (d) detail these anisotropies with an "8" shape; the effective mass along the armchair direction is about an order of magnitude smaller than that along the zigzag direction.

The electrical conductance in monolayer phosphorene is
$$\sigma = (\mu_e e n + \mu_h e p)/L_z \qquad (1)$$
where $e$ is the magnitude of the electron's charge, $\mu_e$ and $\mu_h$ are the mobilities of the electrons and holes, respectively. $n$ and $p$ are the electron and hole densities obtained from the Dirac-Fermi distribution for a given Fermi level (chemical potential) [22] (details in supplementary documents). The vertical thickness for monolayer, $L_z$, is taken to be 0.55 *nm*, which is the average interlayer distance of multilayer phosphorene [14,18,19]. The carrier mobilities are calculated by Bardeen and Shockley's theory [23] of considering the coupling between free carriers and acoustic phonons for this non-polar crystal (more details are provided in the supplementary document). At room temperature, the calculated electron and hole mobilities along the armchair direction are 1800 cm$^2$/v·s and 2300 cm$^2$/v·s, respectively. Recent experimental results support this simple model calculation, finding that the holes are the more mobile of the carriers and reporting the hole mobility around 1000 cm$^2$/v·s [14]. In practice, an improved sample quality will bolster the measured mobility and yield results that are in better agreement with the calculated ones.

The electrical conductance is presented in Figure 2 (e). It depends exponentially on the Fermi level, because of the nearly-constant mobility of the free carriers in the Drude model. In accordance with the above discussion of the effective masses, the conductance is spatially anisotropic (Figures 2 (c) & (d)); the conductance along the armchair direction is about an order of magnitude larger than it is along the zigzag direction, across all doping levels. For a *p*-doping density of 2 x $10^{16}$ m$^{-2}$, the conductance along the armchair and zigzag directions is calculated to be 1.33 x $10^6$ S/m and 1.04 x $10^5$ S/m, respectively.

*Anisotropic thermal conductance*: The thermal conduction in most semiconductors is mediated chiefly by lattice vibrations (phonons) [24-26]; free carriers seldom play important roles in thermal transport in these gapped materials except for heavy-doping cases. Given this approximation, the thermal conductivity $\kappa$ of phosphorene at a finite temperature $T$ can be considered to be comprised of the phonon mode contributions: $\kappa = \sum_\lambda K_\lambda$, where the sum is taken over all phonon branches. The contribution from each mode is given by [27,28]:

$$\kappa_\lambda = \frac{1}{L_x L_y L_z} \sum_{\lambda, q} \tau_\lambda(\vec{q}) C_{ph}(\omega_\lambda)[v_\lambda(\vec{q}) \bullet \hat{n}]^2 , \qquad (2)$$

where $\tau_\lambda$ is the phonon relaxation time, $C_{ph}(\omega_\lambda)$ is the specific heat contribution of the mode, and $v_\lambda(\vec{q})$ is the phonon sound velocity with wave vector . $\hat{n}$ is the unit vector that points along the thermal gradient $\nabla T$.

The phonon dispersion of monolayer phosphorene, as calculated by first-principles DFPT calculations, is detailed in Figure 3 (a). Optical phonons have smaller sound speed ($v_\lambda(\vec{q})$) and much smaller contributions to the heat capacity ($C_{ph}(\omega_\lambda)$ (Details can be found in the supplementary document). Therefore, according to Eq. 2, we neglect those optical-phonon contributions to the thermal conductance, which is similar to the simplification employed in the case of graphene. [28]

Two in-plane acoustic modes, the transverse acoustic mode (TA), and the longitudinal acoustic mode (LA) all exhibit linear dispersions, while the off-plane acoustic mode (ZA) exhibits a parabolic dispersion – in accordance with phosphorene's 2D geometry. The LA phonon stands out for exhibiting an anisotropic speed of sound. The lattice structure of phosphorene decides that the sound speed of the LA branch is larger along the zigzag direction than it is along the armchair direction. The longitudinal vibrations can easily distort the puckered armchair structure, whereas the zigzag lattice is more robust and more difficult to excite. Therefore, the LA branch's speed of sound along the zigzag direction is 8397 m/s, which is almost twice the speed along the armchair direction 4245.5 m/s. According to Eq. 2, the thermal conductance is proportional to the squares of these speeds. Phosphorene's preferred direction of thermal conduction will thus be enhanced along the zigzag lattice direction, orthogonal to the preferred direction of electrical conduction.

To quantify the thermal conductance, the relaxation time ($\tau_\lambda$) of a specific phonon mode ($\lambda$) is calculated by considering nonlinear lattice vibrations that are chiefly mediated by phonon-phonon scattering in the lowest-order approximation[28, 29] is given by:

$$\tau_\lambda = \frac{\omega_m}{\omega^2} \frac{1}{2\gamma_\lambda^2} \frac{Mv^2}{k_B T}, \qquad (3)$$

in which $M$ is the atomic mass, $\omega_m$ is the Debye frequency, $k_B$ is Boltzmann constant, and $\gamma_\lambda$ is the Grüneisen parameter, which is given by the negative logarithmic derivative of the phonon frequency – with respect to the change in volume [27-29]. The average velocity $v$ for a 2D phonon comprised of the LA and TA modes is approximated by $\frac{2}{v^2} = \frac{1}{v_{LA}^2} + \frac{1}{v_{TA}^2}$ in the long wavelength limit [28,29].

An outstanding issue arises when considering only the above two-phonon scattering mechanism: Eq. 3 predicts diverging acoustic phonon lifetimes at the long-wave limit. It is necessary to include three-phonon Umklapp scatterings or scattering with rough boundaries when ω approaches zero. In practice, a truncation scheme is typically implemented in which the long wavelength lifetime is estimated from experimental data.[30] Because there is no such data for monolayer phosphorene, the relevant thermal conductance value used in the present calculation is obtained from data regarding bulk black phosphorous (60 ps). Additional details are provided in the Supplementary Document. It is assumed that the weak van der Waals interactions between phosphorene layers will not dramatically change the long wavelength phonon scatterings within a single layer. For the sake of being systematic, several realistic long-wave lifetimes are used to calculate thermal conductances; it is found that the final results are robust in this regard.

Given the above calculations and approximations, the lattice thermal conductances are plotted versus temperature in Figure 3 (b). The multiple curves, corresponding to different long-wave lifetimes, are plotted. The thermal conductance is consistently larger along the zigzag direction than it is along the armchair direction. That is, the preferred direction for thermal conductance in monolayer phosphorene lies perpendicular to the direction of preferred electrical conductance, as presented in Figure 1.

*Seebeck coefficient*: The Seebeck coefficient (*S*) of monolayer phosphorene is calculated using BTE (see the Supplementary Document for details). Figures 4 (a) and (b) show the *S* for monolayer phosphorene at room temperature and 500K with different free carrier doping densities. The coefficient varies dramatically with doping density, indicating that optimizing the density is crucial for achieving efficient thermoelectric performance. Unlike many of the other features of phosphorene, the Seebeck coefficient is nearly isotropic. These calculated values are comparable with typical thermoelectric materials[8], and are in accord with recent studies of phosphorene [31].

*Thermoelectric figure of merit (ZT):* Figure 4 (c) depicts the thermoelectric figure of merit as a function of free-carrier density along the zigzag and armchair lattice directions, respectively, at room temperature. As expected, given the above observations and discussions, *ZT* exhibits a strong spatial anisotropy in monolayer phosphorene. The armchair direction fosters a larger *ZT* than does the zigzag direction. Given the orthogonal preferred directions of the electrical and thermal conductances, the values of *ZT* along the armchair direction can be enhanced well above the figures of merit of typical materials [3-8]. A doping density of ~ 2 x $10^{16}$ m$^{-2}$ yields a figure of merit well above 1 for reasonable relaxation time (60 ps), which is extremely attractive for applications. For a higher temperature (*T*=500*K*), the figure of merit is even better as shown in Figure 4 (d); it can even reach 2.5 under the same doping density.

Phosphorene's exceptional *ZT* is superior to most third-generation bulk thermoelectric materials under development.[6] According to the Carnot efficiency for heat conversion, phosphorene based thermoelectric devices are poised to reach an energy conversion efficiency of 15~20%,[24] surpassing the criterion for commercial deployment. Outstandingly, unlike most thermoelectric materials whose *ZT* are meager at lower temperatures, the *ZT* of phosphorene is still substantial at room temperature. This feature, plus phosphorene's environment-friendly composition and exceptional mechanical elastic properties, distinguishes it as an exceptional candidate for thermoelectric applications at room temperature.

It is imperative to consider factors that may, in practice, affect the value of *ZT* in phosphorene. First, the role of doping free carriers in driving thermal conductance has not been considered. Recent studies of electron-induced thermal conductance shows that a *p*-doping density of 6 x $10^{16}$ m$^{-2}$ would lead to a thermal conductance around 1.5 W/K·m.[31] This is an order of magnitude smaller than the predicted conductance along the armchair direction (11.8 W/K·m), as seen in Figure 3 (b). Therefore, the contributions of the dopants to the thermal conductance will not significantly modify *ZT*. Second, the calculations and simulations pertain to suspended monolayer phosphorene. Realistically, defects will act to reduce the conductance and diminish the anisotropies. Therefore, modulation doping of 2D structures based on charge transfer [32,33] may avoid scattering defects and preserve the high *ZT*. Meanwhile, substrates will also increase the electronic screening, reducing the band gap and impacting *ZT* slightly. This band gap effect is discussed in the Supplementary Document.

*Conclusion*: Monolayer phosphorene was assessed for its intrinsic thermoelectric capabilities. Accordingly, the electrical conductance, thermal conductance, and Seebeck coefficient of were calculated and used to determine *ZT*. It was found that the electrical and thermal conductances exhibit strong spatial anisotropies such that their respective preferred directions of conductance are orthogonal to one another, resulting in an anisotropic thermoelectric figure of merit that is large along the armchair direction. Thermoelectric figures of merit of 1 and 2.5 can be achieved at exceptionally low temperatures – 300 K and 500 K – using moderate doping. Monolayer phosphorene is thus an attractive candidate for energy applications based on inexpensive and flexible thermoelectric materials that can operate near room temperature.

Method Summary

In present calculations, the atomic structure and electronic structure are calculated by density functional theory (DFT) with the Perdew, Burke, and Ernzerhof (PBE) functional [34]. The Quantum Espresso computing package is used to solve the Kohn-Sham equations, producing the electronic band structure [35]. The band gap is then refined to account for many-body interactions, using the GW approximation via BerkeleyGW.[36] The phonon and thermal conductance calculations are performed in the framework of density functional perturbation theory (DFPT) [37]. Finally, the Seebeck coefficient is calculated using the BTE implemented in the BolzTrap package [38].

Acknowledgement

We acknowledge fruitful discussions with Vy Tran and Anders Carlsson. This work is supported by the National Science Foundation Grant No. DMR-1207141. The computational resources have been provided by the Lonestar of Teragrid at the Texas Advanced Computing Center (TACC).

**References:**

1. Snyder, G. J., &Toberer, E. S. Complex thermoelectric materials. *Nature Mater.* **7**, 105-114 (2008).

2. Curtarolo, S., Hart, G. L. W., Nardelli, M. B., et al. The high-throughput highway to computational materials design. *Nature materials* **12**, 191-201 (2013).

3. Hsu, K. F. *et al*. Cubic AgPb$_m$SbTe$_{2+m}$: bulk thermoelectric materials with high figure of merit. *Science* **303**, 818-821 (2004).

4. Biswas, K., He, J., Blum, I. D., et al. High-performance bulk thermoelectrics with all-scale hierarchical architectures. *Nature* **489**, 414-418 (2012).

5. Rhyee, J. S., Lee, K. H., Lee, S. M., et al. Peierls distortion as a route to high thermoelectric performance in In$_4$Se$_{3-\delta}$ crystals. *Nature* **459**, 965-968 (2009).

6. Zhao, L. D., Dravid, V. P., Kanatzidis, M. G. The panoscopic approach to high performance thermoelectrics. *Energy & Environmental Science* **7**, 251-268 (2014).

7. Hochbaum, A. I., Chen, R., Delgado, R. D., *et al*. Enhanced thermoelectric performance of rough silicon nanowires. *Nature* **451**,163-167 (2008).

8. Zhao, L. D., Lo, S. H., Zhang, Y., *et al*. Ultralow thermal conductivity and high thermoelectric figure of merit in SnSe crystals. *Nature* **508**,373-377 (2014).

9. Balandin, A. A. Thermal properties of graphene and nanostructured carbon materials. *Nature Mater.* **10**, 569-581 (2011).

10. Balandin, A. A., Ghosh, S., Bao, W., *et al*. Superior thermal conductivity of single-layer graphene. *Nano Letters* **8**, 902-907 (2008).

11. Ghosh, S., Bao, W., Nika, D. L., *et al*. Dimensional crossover of thermal transport in few-layer graphene. *Nature Mater.* **9**, 555-558 (2010).

12. Jo, I., Pettes, M. T., Kim, J., *et al.* Thermal conductivity and phonon transport in suspended few-layer hexagonal boron nitride. *Nano Letters* **13**, 550-554 (2013).

13. Sreeprasad, T. S., Nguyen, P., *et al*. Controlled, Defect-Guided, Metal-Nanoparticle Incorporation onto MoS$_2$ via Chemical and Microwave Routes: Electrical, Thermal, and Structural Properties. *Nano Letters* **13**, 4434-4441 (2013).

14. Li, L., Yu, Y., Ye, G. J., *et al*. Black phosphorus field-effect transistors.. *Nature Nanotechnol.* DOI: 10.1038/nnano.2014.35 (2014).

15. Liu, H., Neal, A. T., Zhu, Z., et al. Phosphorene: An Unexplored 2D Semiconductor with a High Hole Mobility. *ACS Nano*, **8**, 4033–4041 (2014).

16. Reich, E. S. Phosphorene excites materials scientists. *Nature* **506**, 19-19 (2014).

17. Koenig, S. P., Doganov, R. A., Schmidt, H., *et al*. Electric field effect in ultrathin black phosphorus. *Appl. Phys. Lett* **104**,103106 (2014).

18. Xia, F., Wang, H., &Jia, Y. Rediscovering Black Phosphorus: A Unique Anisotropic 2D Material for Optoelectronics and Electronics. arXiv:1402.0270 (2014).

19. Fei, R., &Yang, L. Strain-Engineering the Anisotropic Electrical Conductance of Few-Layer Black Phosphorus. *Nano Lett.* DOI: 10.1021/nl500935z (2014)

20. Tran, V., Soklaski, R., Liang, Y., *et al*. Layer-Controlled Band Gap and Anisotropic Excitons in Phosphorene. arXiv:1402.4192 (2014).

21. Allen, P.B., Boltzmann theory and resistivity of metals, in: J.R. Chelikowsky, S.G. Louie (Eds.), *Quantum Theory of Real Materials, Kluwer*, Boston, 219–250 (1996) .

22. Ashcroft, N.W., Mermin, N. D. Solid state physics. Harcourt, Inc., (1976).

23. Bardeen, J. and Shockley W. Deformation Potentials and Mobilities in Non-Polar Crystals. *Physical Review* **80**, 72 (1950).

24. Heremans, J. P., Dresselhaus, M. S., Bel,l L. E., et al. When thermoelectrics reached the nanoscale. *Nature nanotechnology* **8**, 471-473 (2013).

25. Ward, A., Broido, D. A., Stewart, D. A., *et al*. Ab initio theory of the lattice thermal conductivity in diamond. *Physical Review B* **80**, 125203 (2009).

26. Garg, J., Bonini, N., Kozinsky, B., *et al*. Role of disorder and anharmonicity in the thermal conductivity of silicon-germanium alloys: A first-principles study. *Physical review letters* **106**, 045901 (2011).

27. Nika, D. L., Pokatilov, E. P., Askerov, A. S., *et al*. Phonon thermal conduction in graphene: Role of Umklapp and edge roughness scattering. *Physical Review B* **79**, 155413 (2009) .

28. Kong, B. D., Paul, S., Nardelli, M. B., *et al*. First-principles analysis of lattice thermal conductivity in monolayer and bilayer graphene. *Physical Review B* **80**, 033406 (2009).

29. Klemens, P. G. &Pedraza, D. F. Thermal Conductivity of Graphite in the Basal Plane *Carbon* **32**, 735-741 (1994).

30. Slack, G. A.. Thermal conductivity of elements with complex lattices: B, P, S. *Physical Review* **139**, A507-A515 (1965)..

31. Lv, H. Y. , Lu, W. J., Shao, D. F., &Sun, Y. P. Large thermoelectric power factors in black phosphorus and phosphorene. arXiv:1404.5171 (2014).

32. Wang,F., Zhang, Y., Tian, C., Girit, C., Zettl, A., Crommie, M., Shen, and Y.R., Gate-Variable Optical Transitions in Graphene, Science, **320**, 5873 (2008).

33. Ghatak S., Nath Pal, A., and Ghosh, A., Nature of Electronic States in Atomically Thin MoS2 Field-Effect Transistors, ACS Nano, 5, 7707 (2011).

34. Perdew, J. P., Burke, K., &Ernzerhof, M. *Phys. Rev. Lett.***77**, 3865-3868 (1996).

35. Giannozzi, P., *et al*. QUANTUM ESPRESSO: a modular and open-source software project for quantum simulations of materials. *J. Phys.: Condens. Matter* **21**, 395502 (2009),.

36. J. Deslippe, et al. *Comput. Phys. Commun.* **183**, 1269 (2012).

37. Baroni, S., de Gironcoli, S., Dal, C. A., *et al*. Phonons and related crystal properties from density-functional perturbation theory. *Reviews of Modern Physics* **73**,515-562 (2001).

38. Madsen, G. K. H., Singh, D. J. BoltzTraP. A code for calculating band-structure dependent quantities. *Computer Physics Communications* **175**, 67-71 (2006).

**Figures:**

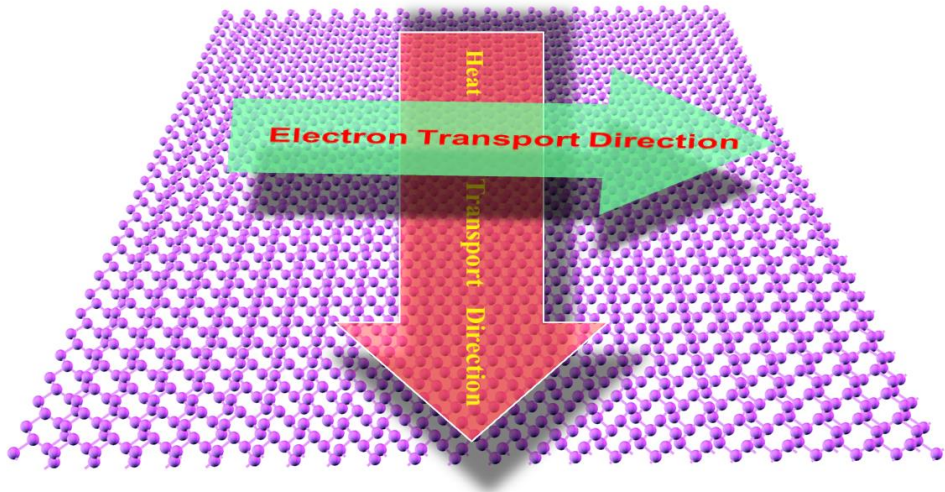

**Figure 1** Schematic of orthogonal electrical conductance and thermal conductance in monolayer phosphorene.

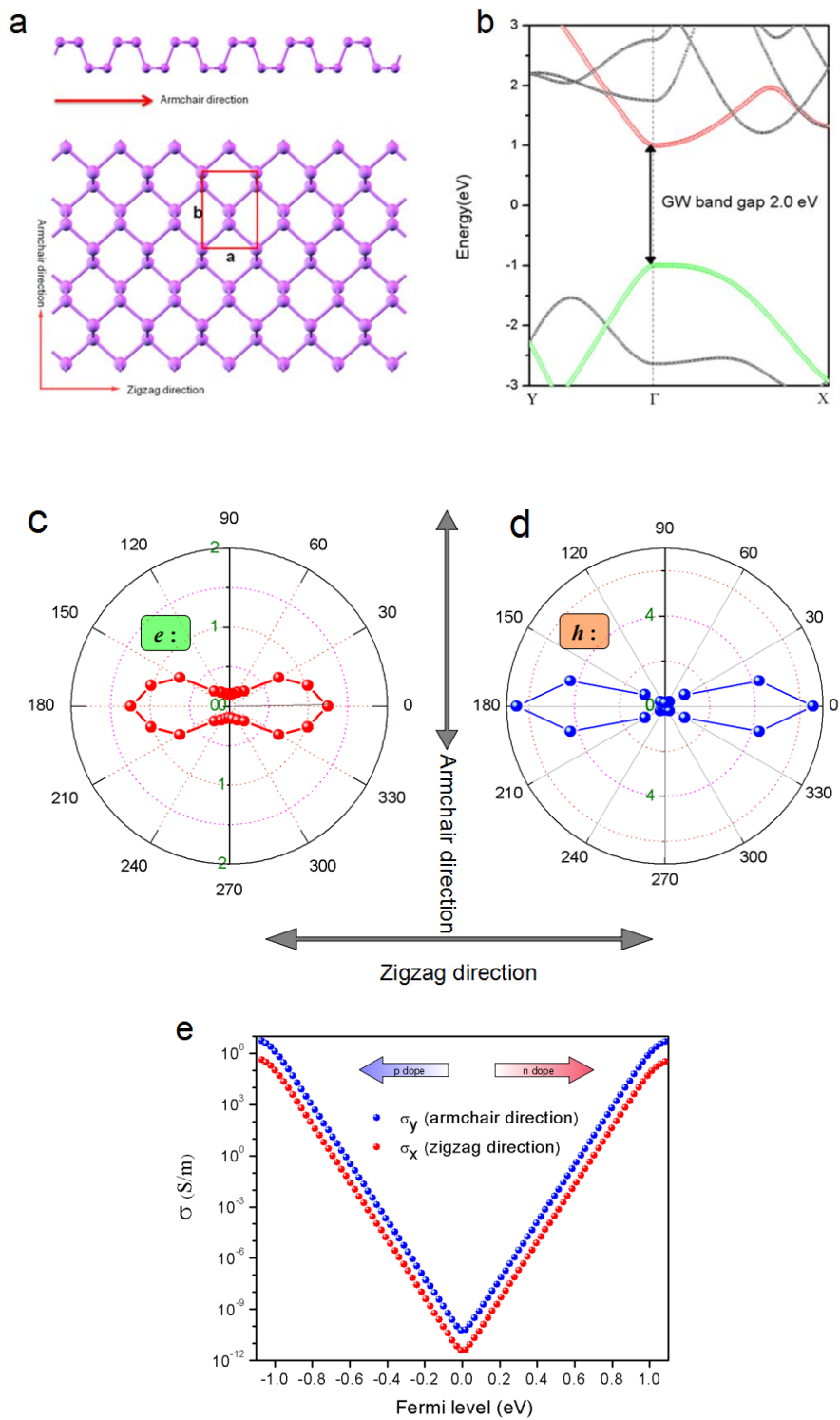

**Figure 2** (a) Atomic structure of monolayer phosphorene. (b) GW-calculated electronic band structure of suspended monolayer phosphorene. (c) and (d) Effective mass of electrons and holes according to spatial directions. (e) Electrical conductance of monolayer phosphorene by the

Drude model. The scale of conductance is a logarithmic scale. The minus sign of the Fermi level means hole *p*-doping and the plus sign means electron *n*-doping.

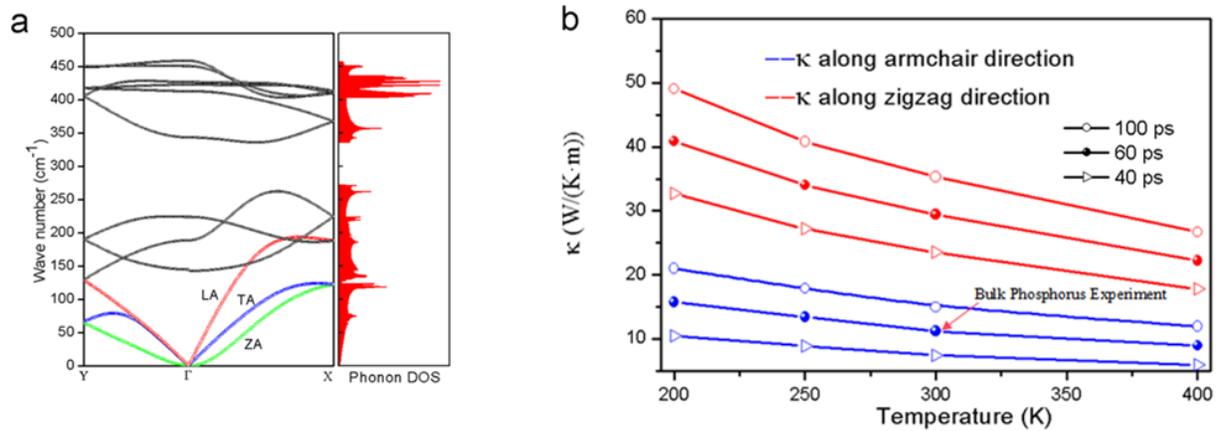

**Figure 3** (a) Phonon dispersion and DOS of monolayer phosphorene. (b) Thermal conductance of monolayer phosphorene. The solid symbols are estimation based on the long-wave relaxation time (60 ps) derived from experimental data of bulk black phosphorus.[30] Different curves are using different long-wave relaxation time.

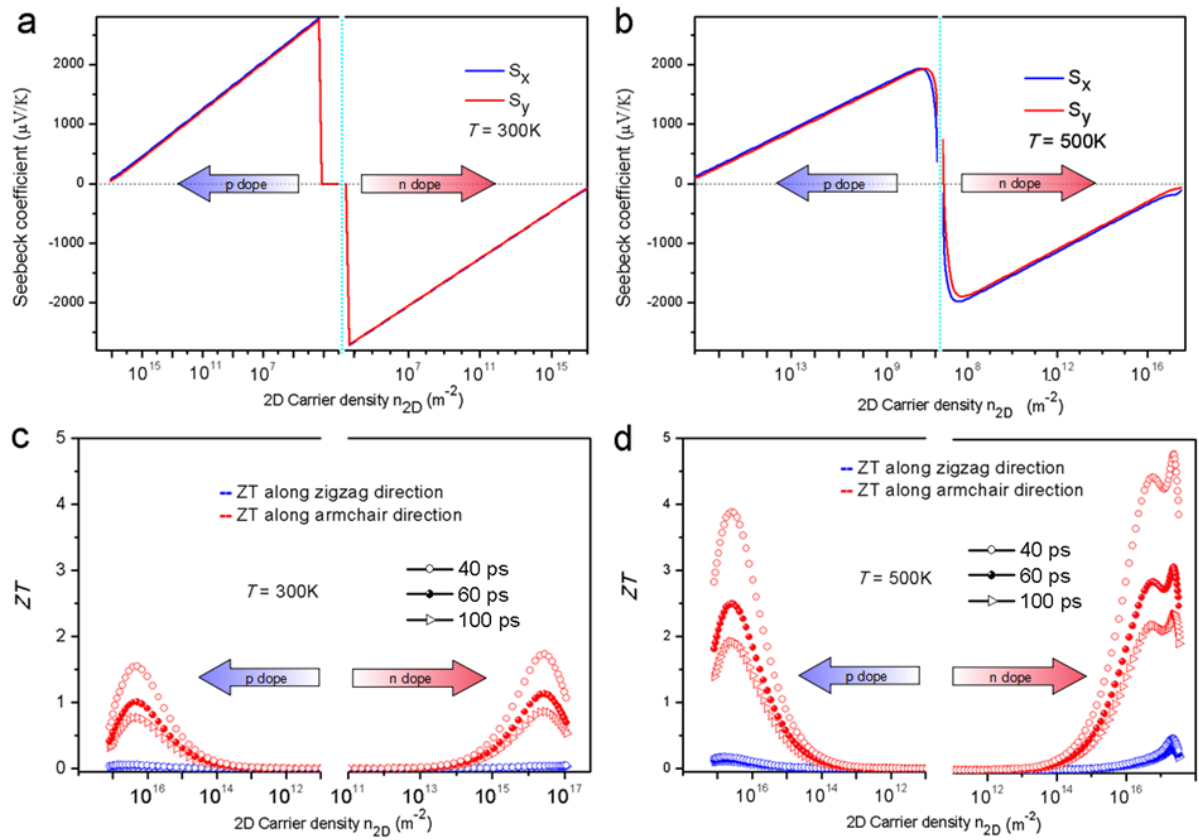

**Figure 4** (a) and (b) Seebeck coefficient according to the doping density, calculated by the BTE at T=300K and 500K, respectively. (c) and (d) The thermoelectric figure of merit according to the doping density under T=300K and 500K, respectively. Different long-wave relaxation time of phonons are included.

# Supplement Information

# Enhanced Thermoelectric Efficiency via Orthogonal Electrical and Thermal Conductances in Phosphorene

Ruixiang Fei, Alireza Faghaninia, Ryan Soklaski, Jia-An Yan, Cynthia Lo, and Li Yang

**Section 1 Electrical Conductance**

The electrical conductance can be calculated by[1]

$$\sigma = (\mu_e en + \mu_h ep)/L_z \quad (S1)$$

where $e$ is the absolute value of an electron's charge, $\mu_e$ and $\mu_h$ are the mobility of electrons and holes. $n$ and $p$ are the electron and hole densities obtained by the Dirac-Fermi distribution for a given Fermi level (chemical potential). The vertical thickness for monolayer phosphorene is $L_z = 0.55$ nm. We will calculate the above carrier densities and mobilities step by step.

*Carrier density:* For 2D semiconductors with quadratic band edges, the corresponding density of states (DOS) is a constant, $\frac{4\pi m^*}{h^2}$, where $h$ is Planck's constant. At a finite temperature ($T$), the carrier density ($n$ and $p$) can be thus calculated by [1]

$$n = \int_{\varepsilon_c}^{\infty} \frac{4\pi m_e^*}{h^2} [\exp(\frac{\varepsilon - \varepsilon_F}{k_B T}) + 1]^{-1} d\varepsilon \quad (S2)$$

$$p = \int_{-\infty}^{\varepsilon_v} \frac{4\pi m_h^*}{h^2} [\exp(-\frac{\varepsilon - \varepsilon_F}{k_B T}) + 1]^{-1} d\varepsilon \quad (S3)$$

Here, $k_B$ is the Boltzmann constant, respectively, and $m^*$ is average effective mass, e.g., $m_e^* = 2[1/m_{ex} + 1/m_{ey}]^{-1}$ and $m_h^* = 2[1/m_{hx} + 1/m_{hy}]^{-1}$. $\varepsilon_F$, $\varepsilon_c$, and $\varepsilon_v$ are the Fermi level, the bottom conductance band energy and the top of valance band energy, respectively.

*Mobility:* Because phosphorene is a non-polar crystal, at a finite temperature (*T*), we calculate the free-carrier mobility using Bardeen and Shockley's theory, [2-4]

$$\mu_{x\_anisotropy} = \frac{eh^3 C_{x\_anisotropy}}{(2\pi)^3 k_B T m_e^* m_d E_{1x}^2} \tag{S4}$$

Here $m_e^*$ is the effective mass along the transport direction, and the density-of-state mass $m_d$ is determined by $m_d = \sqrt{m_{ex}^* m_{ey}^*}$. The deformation potential constant $E_{1x} = \Delta E / (\Delta l_x / l_x)$ is obtained by varying the lattice constant along the transport direction $\Delta l_x$ ($l_x$ is the lattice constant along the transport direction, here is *x* direction) and checking the change of band energy under the lattice compression and strain. The elastic module $C_{x\_anisotropy}$ is obtained via $C_{x\_anisotropy}(\Delta l_x / l_x)^2 / 2 = (E - E_0) / S_0$, where $E - E_0$ is obtained by varying the lattice constant by small amount ($\Delta l_x / l_x \sim 0.5\%$) to obtain the change of the total energy, and $S_0$ is the lattice area in the *xy* plane.

**Section 2 Grüneisen parameter, phonon lifetime and heat capacity**

*Grüneisen parameter:* For a specific branch (λ) of phonons, the corresponding Grüneisen parameter reflecting anharmonic effects is defined as [5,6]

$$\gamma_\lambda = -\frac{A}{\omega_\lambda} \frac{\partial \omega_\lambda}{\partial A} \tag{S5}$$

where *A* is the area of a unit cell of 2D phosphorene, $\omega_\lambda$ is the angular frequency of a specific phonon (λ). In practice, to obtain reliable Grüneisen parameters, we compress and stretch the unit cell and choose averaged values for canceling the first-order errors. As shown in Figures S1 (a) to (c), we have checked the different amounts (±2% and ±4%) of perturbations for calculating Grüneisen parameters of three acoustic phonons. In particular, the Grüneisen parameter of the ZA mode diverges when its momentum approaches zero, as shown in Figure S2 (a). This is due to the unique quadratic band dispersion of the ZA mode, which was also observed in graphene.[6]

*Phonon lifetime:* For a specific phonon branch (λ), its lifetime can be estimated by [6,7]

$$\tau_\lambda = \frac{\omega_m}{\omega^2} \frac{1}{2\gamma_\lambda^2} \frac{Mv^2}{k_B T}, \tag{S6}$$

where $M$ is the atomic mass, $\omega_m$ is the Debye frequency, $k_B$ is Boltzmann constant, and $\gamma_\lambda$ is the Grüneisen parameters. Because of the divergence of the Grüneisen parameter of the ZA mode (shown in Figure S2 (a)), its corresponding lifetime will be extremely small and we can neglect contributions from the ZA mode in the following calculations. Therefore, for simplicity, the average sound speed $v$ for a 2D phonons consisting of LA and TA modes is approximated by the relation $\frac{2}{v^2} = \frac{1}{v_{LA}^2} + \frac{1}{v_{TA}^2}$ in the long-wave length limit.

A particularly important issue of Eq. S6 is that at the long-wave limit of acoustic phonons (the frequency (ω) approaches to zero), the lifetime will diverge as shown in Figure S1 (d). In practice, we have to use a truncation, which is usually based on experimental data.[8] Here we align our results to experimentally measured thermal conductance of bulk phosphorus (11.8 W/K·m) at room temperature in a self-consistent way. We tune the long-wave life time (τ), and insert it into the integral for the thermal conductance (Eq. 2 in the main manuscript) along the armchair direction, which is the poor thermal conducting direction. We stop this iteration once the thermal conductance along the armchair direction of phosphorene is obtained the same as the bulk value. This procedure will give a larger thermal conductance along the armchair direction and the lower limit of *ZT*. This makes our prediction a conservative one. Otherwise, if we fit the bulk value to the thermal conductance along the zigzag direction, that along the armchair direction will be much smaller, giving the upper limit of *ZT*.

*Heat capacity:* For a specific phonon (λ), its heat capacity can be calculated by [6]

$$C_{ph}(\omega_\lambda) = k_B \left(\frac{\hbar\omega_\lambda}{2\pi k_B T}\right)^2 \frac{\exp\left(\frac{\hbar\omega_\lambda}{2\pi k_B T}\right)}{\left[\exp\left(\frac{\hbar\omega_\lambda}{2\pi k_B T}\right) - 1\right]^2} \qquad (S7)$$

The calculated heat capacities of the TA and LA modes are presented in Figure S 2 (e).

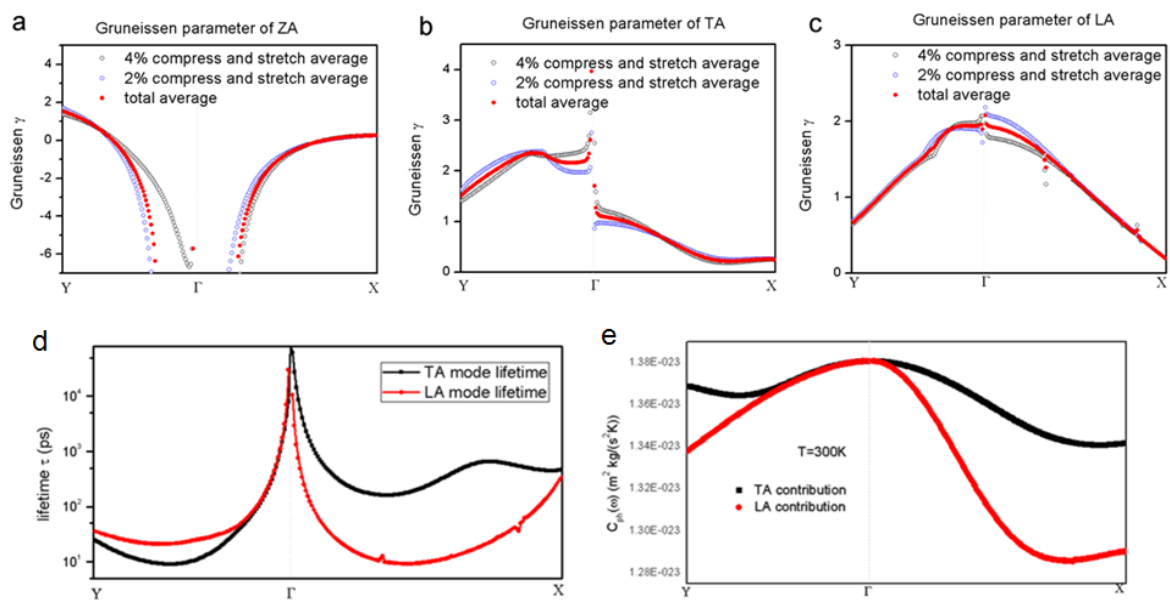

**Figure S1** The Grüneisen parameters of ZA (a), TA(b) and LA (c) phonon modes under different perturbations (± 2% and 4%). (d) Phonon lifetime of LA and TA modes. (e) Heat capacity of LA and TA modes.

**Section 3 Seebeck coefficient**

We use the relaxation time approximation to the Boltzmann transport theory (BTE) to calculate the Seebeck coefficient *S*. This is implemented in BoltzTraP code[9] that uses Fourier interpolation of the calculated energy bands. We derive the following semi-classic nests equations that enable us to determine the Seebeck coefficient S.

$$\sigma_{\alpha\beta}(i,k) = e^2 \tau_{i,k} v_\alpha(i,k) v_\beta(i,k) \tag{S8}$$

$$\sigma_{\alpha\beta}(\varepsilon) = \frac{1}{N} \sum_{i,k} \sigma_{\alpha\beta}(i,k) \frac{\delta(\varepsilon - \varepsilon_{i,k})}{d\varepsilon} \tag{S9}$$

$$\sigma_{\alpha\beta}(T, \varepsilon_F) = \frac{1}{\Omega} \int \sigma_{\alpha\beta}(\varepsilon) \left[ -\frac{\partial f_{\varepsilon_F}(T, \varepsilon)}{\partial \varepsilon} \right] d\varepsilon \tag{S10}$$

$$v_{\alpha\beta}(T, \varepsilon_F) = \frac{1}{eT\Omega} \int (\varepsilon - \varepsilon_F) \sigma_{\alpha\beta}(\varepsilon) \left[ -\frac{\partial f_{\varepsilon_F}(T, \varepsilon)}{\partial \varepsilon} \right] d\varepsilon \tag{S11}$$

$$S = \frac{v_{\alpha\beta}(T, \varepsilon_F)}{\sigma_{\alpha\beta}(T, \varepsilon_F)} \tag{S12}$$

Here, (*i*, *k*) denotes the eigenvalue-band index, and *k* is wave vector. α and β denote the directional coordinates. *N* is the number of *k* points, δ is the unit Kronecker delta function, Ω is the volume of the primitive cell, $\varepsilon_F$ is Fermi level, and $f_{\varepsilon_F}(T,\varepsilon)$ is the Fermi-Dirac distribution. In BTE calculations, we assume that the relax time $\tau_{i,k}$ is direction independent and constant, that is $\tau_{i,k} = \tau$.

**Section 4 Band-gap effects on thermoelectric efficiency**

In main manuscript, we use the GW-calculated band gap for intrinsic and suspended monolayer phosphorene. However, in realistic experimental conditions, samples will be inevitably doped or on substrates. Therefore, the electronic screening is substantially weaker than in the intrinsic case. Consequently, the band gap will be smaller.[10] In order to consider these extrinsic factors, we calculate the thermoelectric efficiency using the DFT band gap, which is a low limit of the band gap.

As shown in Figure S2 (a), the DFT-calculated band structure is similar to the GW-corrected one and the main difference is the DFT band gap is smaller (around 0.92 eV). The figure of merit (ZT) is presented in Figure S2 (b). It is slightly better than the GW result in Figure 4 (c) of the main manuscript at room temperature because the smaller band gap can contribute slightly more free carriers at room temperature. In conclusion, the band gap variation will not substantially change the thermoelectric performance.

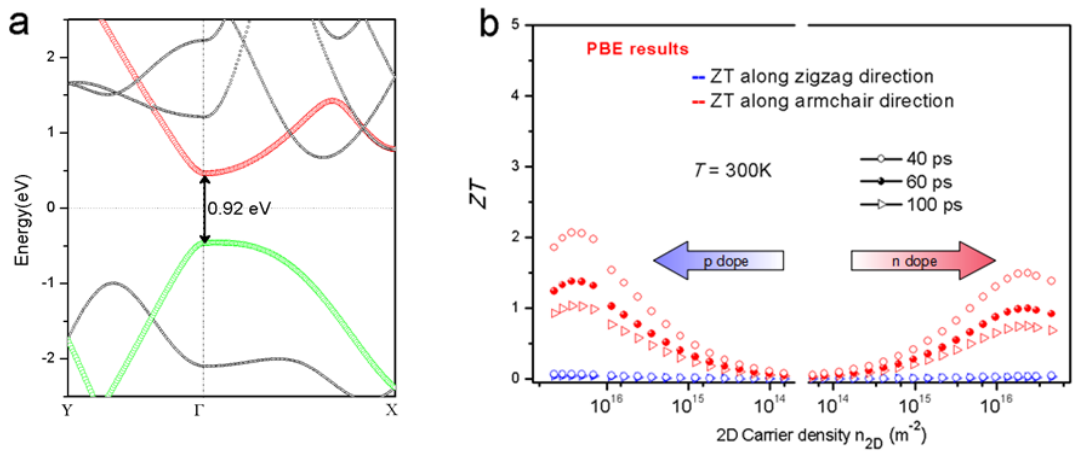

**Figure S2.** The DFT/PBE band structure of phosphorene (a) and the figure of merit (ZT) of phsphorene at room temperature ($T = 300K$) (b).

References:

1. Ashcroft, N.W., Mermin, N. D. *Solid state physics. Harcourt, Inc.*, (1976).

2. Bruzzone, S. & Fiori, G. Ab-initio simulations of deformation potentials and electronmobility in chemicallymodified graphene and two-dimensional hexagonal boron-nitride. *Appl. Phys. Lett.* **99**, 222108 (2011).

3. Takagi, S. i., Toriumi, A., Iwase, M.&Tango,H. On the universality of inversion layer mobility in Si MOSFET's: Part I-effects of substrate impurity concentration. *IEEE Trans. Electr. Dev.* **41**, 2357-2362 (1994).

4. Bardeen, J. and Shockley W. Deformation Potentials and Mobilities in Non-Polar Crystals. *Physical Review* **80**, 72 (1950).

5. Nika, D. L., Pokatilov, E. P., Askerov, A. S., *et al.* Phonon thermal conduction in graphene: Role of Umklapp and edge roughness scattering. *Physical Review B* **79**, 155413 (2009).

6. Kong, B. D., Paul, S., Nardelli, M. B., *et al*. First-principles analysis of lattice thermal conductivity in monolayer and bilayer graphene. *Physical Review B* **80**, 033406 (2009).

7. Klemens, P. G. &Pedraza, D. F. Thermal Conductivity of Graphite in the Basal Plane *Carbon* **32**, 735-741 (1994).

8. Slack, G. A.. Thermal conductivity of elements with complex lattices: B, P, S. *Physical Review* **139**, A507-A515 (1965).

9. Madsen, G. K. H., Singh, D. J. BoltzTraP. A code for calculating band-structure dependent quantities. *Computer Physics Communications* **175**, 67-71 (2006).

10. Hybertsen, M. S., Louie, S. G. Electron correlation in semiconductors and insulators: Band gaps and quasiparticle energies. *Physical Review B* **34**, 5390 (1986).